\documentclass[11pt,a4paper]{article}

\usepackage[left=0.7cm, right=0.7cm, top=0.7cm, bottom=0.7cm]{geometry}

\usepackage{authblk}

\usepackage{fullpage}
\usepackage[utf8]{inputenc}
\usepackage{url}
\usepackage[noend]{algpseudocode}
\usepackage{verbatim}
\usepackage{amsmath, amssymb}
\usepackage{physics}
\usepackage[pdftex]{graphicx}
\usepackage{booktabs}
\usepackage{float}
\usepackage{array}
\usepackage{caption}
\usepackage{subcaption}
\usepackage{setspace}
\usepackage{tabto}
\usepackage{lineno}
\usepackage{bbold}
\usepackage[shortlabels]{enumitem}
\usepackage{color}
\usepackage{soul} 

\usepackage{multicol,multirow}
\usepackage{amsmath,amssymb,amsfonts}
\usepackage{mathrsfs}
\usepackage{amsthm}
\usepackage{apacite}
\usepackage{rotating}
\usepackage{appendix}
\usepackage[authoryear]{natbib}
\usepackage{ifpdf}
\usepackage[T1]{fontenc}%
\usepackage{times}%
\usepackage{sourcesanspro}%
\usepackage{newtxmath}%
\usepackage{textcomp}%
\usepackage{xcolor}
\usepackage{tikz}
\usetikzlibrary{bayesnet}
\usetikzlibrary{arrows,intersections,arrows.meta}
\usetikzlibrary{graphs}
\usepackage{color}
\usepackage{graphicx}
\newcommand{\Ray}{3.8cm}

\usepackage{microtype}

\newcommand*\patchAmsMathEnvironmentForLineno[1]{%
      \expandafter\let\csname old#1\expandafter\endcsname\csname #1\endcsname
      \expandafter\let\csname oldend#1\expandafter\endcsname\csname end#1\endcsname
      \renewenvironment{#1}%
         {\linenomath\csname old#1\endcsname}%
         {\csname oldend#1\endcsname\endlinenomath}}%
    \newcommand*\patchBothAmsMathEnvironmentsForLineno[1]{%
      \patchAmsMathEnvironmentForLineno{#1}%
      \patchAmsMathEnvironmentForLineno{#1*}}%
    \AtBeginDocument{%
    \patchBothAmsMathEnvironmentsForLineno{equation}%
    \patchBothAmsMathEnvironmentsForLineno{align}%
    \patchBothAmsMathEnvironmentsForLineno{flalign}%
    \patchBothAmsMathEnvironmentsForLineno{alignat}%
    \patchBothAmsMathEnvironmentsForLineno{gather}%
    \patchBothAmsMathEnvironmentsForLineno{multline}%
    }
    \include{def}
\def\dispmuskip{\thinmuskip= 3mu plus 0mu minus 2mu \medmuskip=  4mu plus 2mu minus 2mu \thickmuskip=5mu plus 5mu minus 2mu}
\def\textmuskip{\thinmuskip= 0mu                    \medmuskip=  1mu plus 1mu minus 1mu \thickmuskip=2mu plus 3mu minus 1mu}
\def\beq{\dispmuskip\begin{equation}}    \def\eeq{\end{equation}\textmuskip}
\def\beqn{\dispmuskip\begin{displaymath}}\def\eeqn{\end{displaymath}\textmuskip}
\def\bea{\dispmuskip\begin{eqnarray}}    \def\eea{\end{eqnarray}\textmuskip}
\def\bean{\dispmuskip\begin{eqnarray*}}  \def\eean{\end{eqnarray*}\textmuskip}

\def\D{{\cal D}}

\def\argmax{\text{\rm argmax}}

\newcommand{\keywords}[1]{\textbf{Keywords:} #1}

\title{Bayesian Adaptive Trials for Social Policy}

\author[1]{Sally Cripps$^*$}
\author[1,2]{Anna Lopatnikova}
\author[1]{Hadi Mohasel Afshar}
\author[3]{Ben Gales}
\author[1]{Roman Marchant}
\author[1]{Gilad Francis}
\author[1]{Catarina Moreira}
\author[4]{Alex Fischer}

\date{}

\affil[1]{Human Technology Institute, University of Technology Sydney, Australia \\\texttt{sally.cripps@uts.edu.au}}
\affil[2]{Discipline of Business Analytics, The University of Sydney, Australia }
\affil[3]{Paul Ramsay Foundation, Australia }
\affil[4]{Australian National University, Australia }

\date{}  

\begin{document}
\maketitle

\abstract{This paper proposes Bayesian Adaptive Trials (BAT) as both an efficient method to conduct trials and a unifying framework for evaluation social policy interventions, addressing limitations inherent in traditional methods such as Randomized Controlled Trials (RCT). Recognizing the crucial need for evidence-based approaches in public policy, the proposal aims to lower barriers to the adoption of evidence-based methods and align evaluation processes more closely with the dynamic nature of policy cycles.  BATs, grounded in decision theory, offer a dynamic, ``learning as we go'' approach, enabling the integration of diverse information types and facilitating a continuous, iterative process of policy evaluation.  BATs' adaptive nature is particularly advantageous in policy settings, allowing for more timely and context-sensitive decisions. Moreover, BATs' ability to value potential future information sources positions it as an optimal strategy for sequential data acquisition during policy implementation. While acknowledging the assumptions and models intrinsic to BATs, such as prior distributions and likelihood functions, the paper argues that these are advantageous for decision-makers in social policy, effectively merging the best features of various methodologies.\\ }

\keywords{impact evaluation; Bayesian adaptive trials; evidence-based policy; research to policy gap; instrumental philanthropy\\}


\noindent
\textbf{Policy. }
There is a crucial need for practical evidence-based approaches in public policy.  This paper proposes Bayesian Adaptive Trials (BAT) as a unifying framework for evaluating social policy interventions, addressing limitations inherent in traditional methods such as Randomized Controlled Trials (RCT). BATs lower barriers to the adoption of evidence-based methods and align evaluation processes more closely with the dynamic nature of policy cycles. BATs, grounded in decision theory, offer a dynamic, ``learning as we go'' approach, enabling the integration of diverse information types and facilitating a continuous, iterative process of policy evaluation.   BATs adaptive nature is particularly advantageous in policy settings, allowing for more timely and context-sensitive decisions. 


\maketitle


\section{Introduction}

This paper proposes the use of Bayesian Adaptive Trials (BAT) as a powerful framework to continually evaluate the impact of policy interventions. Most people acknowledge that a rigorous evidence-based approach to public and social policy and associated expenditure is crucial for effective resource allocation. It helps identify the most effective programs and initiatives, ensuring that limited public resources are invested where they yield the most significant benefit.  In the absence of price signals this is key to support productivity improvement and efficacy. Yet most would also acknowledge that governments and the social sector more generally have had very limited success in this area \citep{banerjee2017proof,gollust2017mutual,williams2020external}. There are several reasons for this; policy decisions are often not evidence-based but are the product of a myriad of factors, - individual personalities, interest groups, and the ideologies of the government of the day, to name a few. On those occasions where there is a willingness to embrace evidence-based approaches, adoption has been slow, impeded by a range of practical problems related to the complexity, diversity, and context-dependence of responses to social programs \citep{allcott2015site,bates2017generalizability,bedecarrats2020randomized, gugerty2018ten}.  

This paper argues that the BAT framework mitigates several of the problems raised by the authors above. BATs lower the barriers to broader adoption of evidenced-based methodologies, and provide a systematic framework for embedding evidenced-based methodologies which are more in tune with the policy cycle.

At their core, BATs are an adaptive  {\it learning as we go approach} to discovering which initiatives work best and under what conditions.  They rely on the principles of Bayesian reasoning to combine various types of information,  acquired either sequentially or simultaneously, in a logically consistent manner.  In addition, BATs are set in a Bayesian decision theoretic framework \citep{muller2006bayesian}, and as such are able to attach value to potential sources of future information, so that the BATs can act as an optimal sequential data acquisition strategy during policy implementation \citep{cripps2023uncertainty,Marchant2014Uai}. This continuous iterative approach to policy evaluation and implementation is a fundamental shift away from the retrospective evaluation of policies as either successes or failures.

Impact evaluation is a hot topic for many governments \citep{leigh2023evaluating}.  The difficulties that attend impact evaluation often stem from the limitations of the prevalent impact evaluation methodologies, such as Randomized Controlled Trials (RCTs), the ``gold standard’’ traditional impact evaluation method. 
The limitations of RCTs include the cost, the length of time, lack of agility to change as circumstances change, and the low success rate \citep{bedecarrats2020randomized}. There are also challenges in replicating and scaling results \citep{williams2020external,epstein2012program}. {\it Innovations for Poverty Action} have conducted many randomized control trials (RCTs) to evaluate policy impact, and while some  found evidence of high impact, most  were inconclusive \citep{liket2014aren,gugerty2018ten}.  Other common methods of impact evaluation, which are more flexible and less costly than RCTs such as quasi experimental methods and expert opinion have other limitations: lack of statistical rigour and difficulties establishing causality.  External validity -- generalizability of findings to populations others than those under study -- has been difficult to establish across many studies \citep{allcott2015site,banerjee2017proof,deaton2020randomization}.  This raises the question of whether the resources allocated to evaluations would be better spent monitoring outcomes and efforts \citep{gugerty2018ten}.

Bayesian Adaptive Trials (BAT) can serve both as an efficient method to conduct trials and as a unifying framework for evaluating social policy interventions. 
BATs combine the best features of many methodologies, but arguably their main advantage is that they seamlessly expand to form a broader system of continuous organisational learning and policy formation. BATs are particularly valuable for pilots with the ultimate goal of scaling in a similar context, as opposed to academic research that seeks to find treatment effects transportable to any context.  When applied pilots, BATs can be the start of a continuous learning process.  While there are assumptions, such as prior distributions, and models of data-generating processes (likelihood functions), we argue that these drawbacks are in fact advantages for decision-makers in social policy.   

Setting impact evaluation systems in a decision-theoretic framework is not new, \citep{wald47,von1947theory}; however, the computational cost of evaluating the high dimensional integrals has limited the domain of its application.  Recent technological \citep{owens2008gpu} and algorithmic advances \citep{luengo2020survey} have enabled the practical implementation of decision-theoretic methods across many fields, such as medicine, marketing, and robotics \citep{schrage2021transformational, slam}.  Recently, adaptive (although not Bayesian) designs have been proposed in the policy context \citep{offer2021adaptive,esposito2022adaptive}. BATs are a prominent form of adaptive clinical trials in medicine because of their flexibility and data efficiency \citep{warner2021low}.   Adaptive trials are increasingly preferred over traditional fixed clinical trial designs \citep{fors2020current,noor2022uptake} because they provide a faster, more flexible, efficient, and ethical way to conduct clinical trials \citep{thorlund2018key,guidance2019adaptive}.  Adaptive platform trials played a crucial role in the research response to COVID-19 \citep{vanderbeek2022implementation}.  In contrast to fixed-design trials, adaptive trials allow for modifications to the trial protocol based on accumulated data. The efficient adaptations are particularly valuable in situations that bear strong similarities to social policy contexts, such as those characterized by high data collection costs, large numbers of potential treatments, response variability across patients,  time constraints, and significant ethical concerns \citep{barker2009spy,kim2011battle,chaudhuri2020bayesian,bedecarrats2020randomized,berry2010bayesian}.   In marketing, ``perpetual'' BATs are widely used in areas such as multiarm A/Bn testing, marketing mix modeling, price discovery, and recommender systems \citep{schrage2021transformational}.

The remainder of the paper is structured in the following way; Section~2 gives an overview of BATs.  Section~3 highlights the utility of BATs as a framework for evidence systems in the social policy setting.  Section~4 demonstrates, using a numerical example, the efficiency of BATs in finding the optimal dose of tutoring hours in a simulated educational setting.  Section 5 discusses the implications of BATs for organisational learning and strategy. Section 6 concludes.

\section{Bayesian Adaptive Trials: An Introduction}
\label{sec:bats}

Bayesian Adaptive Trials (BATs) represent an experimental approach aimed at establishing causal relationships between test treatments and their effects.  Similarly to RCTs, BATs employ random assignment of participants to treatment and control groups of the trial to reduce the influence of factors not directly under experimental control.  Unique to BATs is their foundation in the Bayesian decision-theoretic framework, increasingly popular in social science \citep{jackman2009bayesian}.  The Bayesian decision-theoretic foundation confers several key benefits. It allows for greater flexibility in trial design, adaptability to evolving data, and a more direct alignment with the information needs of decision-makers.  Importantly, these advantages are achieved without compromising the experimental rigour essential for reliable causal inference.

In this section, we present a description of the algorithmically {\it greedy} version \citep{cormen2022introduction} of the decision-theoretic framework for the applications of BATs to social policy, for a more complete discussion see \cite{muller2006bayesian} and \cite{cressie2023decisions}. We note that a decision-theoretic framework is fundamentally different from an RCT framework and we use a simple example to draw analogies between the two frameworks as well as highlight differences.  

Suppose a government has the goal of improving educational outcomes for children experiencing social disadvantage.  To use the language of experimental design, the children from socially disadvantaged backgrounds are the {\it target population}. The issue, which needs to be established at the outset, is what success looks like.  Does it look like a higher percentage of the target population will complete high school?  Or that the difference in school grades between the target population and others will be reduced?  Or that by the age of 25 say, the number of children  \emph{Not-in-Education, Employment or Training  (NEET)} will have declined?  All these are questions concerning values we attach to outcomes, otherwise known as utilities, which we will denote by $u$.  Utilities are not known precisely, they are random variables because they depend upon future outcomes and unknown quantities.

The next issue concerns the identification of which actions, denoted by $a$, are available to decision-makers.  Again in the language of experimental design, we can view these actions as {\it treatments}. For example, the introduction of a school mentorship program or an after-school tutoring program are treatments that will impact utilities. Conditional on the initiative one needs to determine how it will be implemented.  Unlike utilities,  actions concerning the choice and implementation of initiatives are not random variables; they are decisions, and in a decision-theoretic framework, they should be made according to some objective function, such as maximum expected utility. 

Suppose the chosen initiative is to provide after school tutoring to selected students.  The treatment or actions that need to be decided upon include what is the impact of  tutoring on student performance?  What is the number of tutoring hours that will give the maximum improvement in grades while containing costs.  These unknown quantities, about which we hope to learn from the data generated by the trial, are sometimes called {\it treatment effects}, and we denote them generically by $\theta$.  In Section~\ref{sec:realexample}, we give an example where  $\theta$ if be the unknown function which relates hours of after-school tutoring to improvement in grades.  But $\theta$ could be something more complex, such as the causal pathways that lead to childhood obesity, \cite{zhu2023}.  

 Finally  BATs need to specify a probabilistic model that connects unknown quantities, $\theta$ to experimental data or outcomes, which we denote by $\mathbf y$.  Our utility is then a function of these three quantities: actions or treatments, $a$, the unknown treatment effects, $\theta$, and the outcomes $\mathbf y$, so we write $u(a,\theta,\mathbf y)$ to show this dependency.  The probabilistic model which relates unknown quantities to outcomes is a fundamental difference between RCTs and BATs.

There are two components in this probabilistic model; the first component is a prior distribution for the unknown quantity $\theta$, denoted by $p(\theta)$. The purpose of this prior distribution is to encode previous knowledge such as results of previous studies on the relationship between the number of hours of tutoring and the improvement in grades, or the opinions of teachers and other experts.  In forming priors it is important to ask such questions as - is there a large variability in opinions or previous research findings? Or is previous research relevant for the current context? Decision-makers are free to choose this prior distribution.   They may choose not to take into account any existing knowledge in which case the prior is known to be uninformative or objective and there is a huge literature on how to form such priors, \cite{berger2015overall}. Alternatively, they may wish to include expert opinion, using, for example, prior elicitation techniques, \cite{falconer2022}.    The key point is that the prior is a mechanism by which other information, opinions or voices can be brought into the initiative and public debate.  Although a prior is subjective, it is explicitly stated and therefore requires decision-makers to justify its choice, as well as enable them, or others, to report on the sensitivity of inference to it.

The second component of the probability model is the likelihood function, denoted by $p(\mathbf y|\theta)$. This is the assumed model for the process that generates outcomes, $\mathbf y$, if the treatment effect, $\theta$, were known.  It is analogous to a {\it theory of change}, \citep{weiss1997theory}.  For example, if we knew the true relationship between the number of hours of tutoring and improvement in grades how {\it likely} are the outcomes we observe from the trial and hence the name likelihood function.  The complexity of the likelihood function will reflect the complexity of the issue at hand.  We give a simple example in Section~\ref{sec:realexample}, but clearly, if we wish to learn the causal pathways by which children succeed at school, and we propose to do this by identifying combinations of causal factors and how they vary over time, then the likelihood model would be very complex, \cite{zhu2023}.

Our initial belief about $\theta$, given by our prior, $p(\theta)$, is updated when we observe data $\mathbf y$ via the likelihood function, $p(\mathbf y|\theta)$, to give a new posterior belief about $\theta$, $p(\theta|\mathbf y)\propto p(\theta)p(\mathbf y|\theta)$.  It is called the posterior belief  because it now depends upon the data we observed in the trial. This posterior belief at time $t$ then becomes the new prior for time $t+1$, and so the process continues enabling  decision-makers or agencies to continually learn from the data generated by the initiative.  It is this process of continually learning that gives BATs their agility.

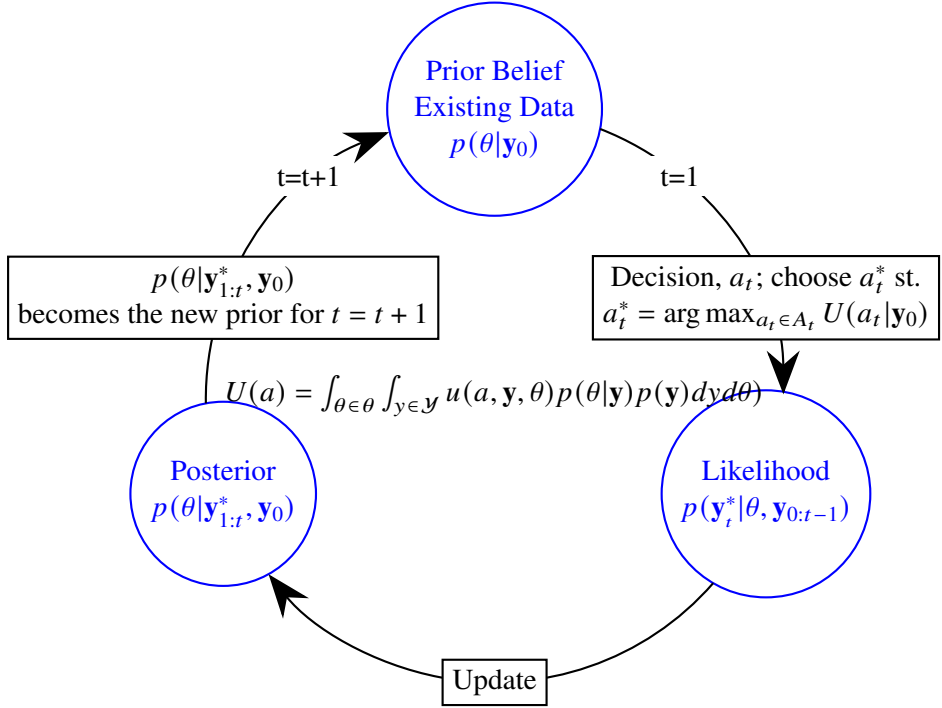
\begin{figure}
\begin{tikzpicture}
\node[circle,blue,minimum width=2cm,minimum height=1cm,draw,thick,name path=n1,align=center,thick] (A) at (90:\Ray) {Prior Belief\\
Existing Data
\\ $p(\theta|\mathbf y_0)$ };
\node[circle,blue, minimum width=2cm,minimum height=1cm,draw,thick,name path=n2,align=center,thick] (B) at (200:\Ray) {Posterior\\$p(\theta|\mathbf y^*_{1:t},\mathbf y_0)$};
\node[circle,blue,minimum width=2cm,minimum height=1cm,draw,thick,name path=n3,align=center,thick] (C) at (-20:\Ray) {Likelihood \\ $p(\mathbf y^*_{_{t}}|\theta,\mathbf y_{0:t-1})$};
\path[name path=c] circle (\Ray);
\path[name intersections={of=n1 and c,name=i1},
      name intersections={of=n2 and c,name=i2},
      name intersections={of=n3 and c,name=i3}
     ];
\begin{scope}[thick]
\pgfsetarrowsend{Stealth[scale=2.5]}
\pgfpathmoveto{\pgfpointanchor{i3-2}{center}}
\pgfpatharcto{\Ray}{\Ray}{0}{0}{0}{\pgfpointanchor{i2-2}{center}}
\pgfusepath{draw}
\pgfpathmoveto{\pgfpointanchor{i1-2}{center}}
\pgfpatharcto{\Ray}{\Ray}{0}{0}{0}{\pgfpointanchor{i3-1}{center}}
\pgfusepath{draw}
\pgfpathmoveto{\pgfpointanchor{i2-1}{center}}
\pgfpatharcto{\Ray}{\Ray}{0}{0}{0}{\pgfpointanchor{i1-1}{center}}
\pgfusepath{draw}
\end{scope}
\node[rectangle,draw, thick,fill=white] at (-90:\Ray) {Update};
\node[rectangle,draw,thick,fill=white,align=center] at (20:\Ray) {Decision, $a_{t}$;
choose $a_t^*$ st. \\$a_t^{*}=\arg\max_{a_t\in A_t}U(a_{t}|\mathbf y_0)$};
\node[rectangle,draw,thick,fill=white,align=center] at (160:\Ray) {$p(\theta|\mathbf y^*_{1:t},\mathbf y_0)$\\becomes the new prior for $t=t+1$};
\node[rectangle,thick] at (-90:0) {$U(a)=\int_{\theta\in\mathcal{\theta}}\int_{y\in \mathcal{Y}}u(a,\mathbf y,\theta)p(\theta|\mathbf y)p(\mathbf y)dyd\theta)$};
\node[rectangle,fill=white] at (130:\Ray) {t=t+1};
\node[rectangle,fill=white] at (50:\Ray) {t=1};
\end{tikzpicture}
\caption{Bayesian Adaptive Trials in a decision-theoretic framework, for initiative $i$.  The quantity $U(a)$, is the expected utility of implementing initiative $a\in A$, where the expectation is w.r.t to the joint distribution $p(\theta,y)=p(\theta|y)p(y)$.}
\label{fig:BAT_dec}
\end{figure}

Figure~\ref{fig:BAT_dec} is a graphical representation of the process for a given initiative.  For any initiative $i$ there is an unknown quantity $\theta_i$, that we wish to learn, and a set of actions, analogous to experimental designs, from which we can choose at times $t=1,\ldots, T$, namely $A_{it}$.  For clarity of exposition we will drop the conditioning on the initiative  $i$, and take it as given.  The initial prior distribution $p(\theta|\mathbf y_0)$ depends upon some existing information, denoted by $\mathbf y_0$.  There are several ways the initiative can proceed, and Figure~\ref{fig:BAT_dec} shows an example of an adaptive trial where actions $a_t\in A_t$, such as the recruitment of schools or individuals, change according to some expected utility, decided prior to the trial,
\begin{equation}
U(a_t)=\int_{\theta\in\mathcal{\theta}}\int_{y\in \mathcal{Y}}u(a_t,\mathbf y_t,\theta|\mathbf y_{1:t-1})p(\theta,\mathbf y_t|\mathbf y_{1:t-1})d\mathbf y_td\theta.
\label{eq:greedy}
\end{equation}
Note that Equation~\eqref{eq:greedy} is a special case of a more general formula proposed by \cite{muller2006bayesian}, where the utility of a sequence of actions is maximized. The specific chosen action, or experimental design,  $a^*_t$ is given by
\[
    a^*_t=\arg\max_{a_t\in A_t}U(a_t).
\]
As such the optimal actions or experimental design may change overtime as we observed data $\mathbf y_{1:t}$ and update our assessment of $p(\theta|\mathbf y_{1:t})$. We discuss this advantage of BATs in Section~\ref{sec:alignment}. The utility function typically includes some measure of learning, for example schools may be recruited into the process by adaptively selecting those which are maximally informative about $\theta$, as measured by mutual information, for example \cite{cripps2023uncertainty}. In addition to a learning criteria the utility function often also includes a criteria which measures benefit, such as the expected improvement in school grades.   Importantly decisions are made based on the expected utility function where the expectation is w.r.t the posterior distribution of $\theta$, and the distributions of future realisations of data.  As in all initiatives, there is often a tension between learning  the effectiveness of an initiative and giving the initiative to those who will most benefit. 

Inference concerning the efficacy of the initiative is made via the posterior distribution $p(\theta|\mathbf y_{1:t})$, and decisions are made by the consequent expected utility $U(a_t)$, which can then be used to compare the performance of different initiatives and any given point in time.  The practical implicatons of this are outlined in Sections~\ref{sec:cost}~and~\ref{sec:alignment}. The stopping criteria for BATs is also given by the utility function, and needs to be defined prior to the start of the initiative.  In the school mentorship program for example the stopping rule could be when the expected benefit is less than the cost of rolling it out to the next school, where the next school is selected based on the criteria above. 

Finally, it is worth noting that the specification of utilities, priors and likelihood functions are seen as drawbacks to BATs.  We argue the reverse.  The process and discussion of forming prior beliefs is a very useful exercise; it involves engaging with the community, listening to experts, and demonstrating to those unfamiliar with the concept of uncertainty, that its quantification is vital to guide the development of the initiative. Similarly, a discussion of utilities connects experts in data science and machine learning to communities with lived experience and policy makers, these ideas are discussed further in Section~\ref{sec:Theoryofchange}. 

\section{BATs for Social Policy Impact Evaluation} 

Impact evaluations are a crucial tool in social policy, used by social sector organisations, government bodies, foundations, and other stakeholders to assess and improve the quality, efficiency, and effectiveness of policies and programs.  Essential for ensuring transparency and productivity improvements through learning feedback loops, they play a significant role in the broader context of evidence-based policymaking.  
The role of impact evaluations is to help determine whether changes in outcomes are due to specific programs, initiatives, or policies by establishing a causal link between the program or policy and the intended outcomes. 
Establishing the causal link requires experimental quantitative data, which needs to be combined with other sources of information to ensure both the quality of the experiment and its policy relevance \citep{gertler2016impact}.

\cite{gugerty2018goldilocks} advocate for the development of \emph{right-fit evidence systems} that embed impact evaluations in a broader evidence system that also includes actionable and timely operational data, such as monitoring data and cost information.  According to \cite{gugerty2018ten}, causal attribution of impact requires a counterfactual, in traditional best practice achieved with an RCT.  Monitoring data and other observational data cannot substitute for randomization in establishing a credible counterfactual.   Nevertheless, these data play a critical role in ensuring the successful implementation and scaling of programs and provide valuable information for the decision-maker when impact evaluations are not suitable or not feasible. A right-fit evidence system would integrate all sources of valuable and actionable data to support decision-makers with policy-relevant insights.

The structure of BATs and their decision-theoretic origin make them an excellent candidate to form a structured framework and a statistical scaffold for the right-fit evidence systems advocated by \cite{gugerty2018goldilocks}.  BATs, like RCTs,  deliver rigorous impact evaluations and are more efficient than RCTs.  However, the main advantage of BATs is that they seamlessly expand to form a broader system of continuous organisational learning (Sections~\ref{sec:bats} and \ref{sec:learning}).   BATs offer several critical benefits: (1) Cost and data efficiency; (2) Alignment with decision-making; (3) Closer alignment with ethical principles of beneficence and equipoise; (4) A path towards personalisation of treatments, supports, and services; and (5) A path toward validating theories of change.

\subsection{Cost and Data Efficiency}
\label{sec:cost}

Government agencies and NGOs operate under significant budgetary constraints, making cost efficiency paramount \citep{thomas2019economic}.  By identifying which initiatives deliver the best outcomes per dollar spent, evaluations enable policymakers to make informed decisions about allocating scarce resources.  Interest in impact evaluations has intensified as governments strive to balance fiscal responsibility with the need to address evolving social, economic, and environmental challenges.   

The cost efficiency of impact evaluations themselves -- the ability to extract the most rigorous information about the causal impact of an initiative at the lowest cost -- is a crucial requirement for a modern evaluation system.  Yet high-quality impact evaluations can be costly, as they require significant managerial, technical, and on-the-ground resources as well as investments in training and information technology infrastructure.  As valuable as impact evaluations can be, they compete for resources with service delivery and implementations of existing initiatives.  As a result, it is essential that the evaluation itself is cost-effective and delivers valuable information that leads toward desired outcomes -- in other words, yields better outcomes per dollar spent than alternative uses of the scarce resources.  If a trial delivers insufficient benefits, the decision not to run it results in significant cost savings.  Current impact evaluation approaches lack a framework to arrive at rigorous answers to the question ``Is running an experiment is a good use of resources?''\footnote{As an example, it is well-understood that quasi-experimental designs and observational studies can be less costly to conduct than RCTs, but also yield a less rigorous assessment of causal treatment effects.  Without a structured quantitative framework, it is difficult to compare which method will yield more valuable information per dollar.}   The decision-theoretic framework of BATs provides the necessary rigour to address this question.    

The decision-theoretic BAT is efficient by design.  BATs' cost efficiency relative to the other evaluation methods stems from multiple sources. First, in contrast to the RCTs, which yield a binary “yes/no” answer regarding a treatment's statistical significance, BATs give detailed and actionable information on the treatment's range of possible outcomes.  Second, BATs provide a rigorous way to include information from multiple sources, such as from prior studies, including quasi-experimental designs and expert interviews; this prevents the loss of information at handoffs from one study to another that often occurs in impact evaluations.  Third, BATs provide a way to adjust the trial based on the accumulated trial data, resulting in scarce resources being deployed in a manner that maximizes information gain.

Consider a typical scenario, where a public agency or charitable foundation funds an RCT-based pilot based on compelling evidence acquired in a different but related context, e.g., in a different geography with similar characteristics.  The pilot demonstrates a reasonable average treatment effect but higher-than-expected response variability.  As a result, the treatment effect fails to breach the statistical significance threshold, and the agency does not adopt the initiative, failing to capitalize on the valuable information the pilot delivered.

A BAT would improve on this outcome in a number of ways.  The Bayesian approach delivers information beyond the binary conclusion that the estimated treatment effect is not statistically significant - it provides actionable information on what the probable range of outcomes might be if the treatment is rolled to further participants – e.g. scaled up to the entire target population.  The value of this information is particularly clear when comparing multiple treatments.  For example, in the context of medical research, a recent paper \citep{lammers2023use} uses the BAT framework to reanalyze the results from a (modified) RCT on optimal platelet and plasma ratios for blood transfusions.  The RCT compared two treatment options and found that the options do not differ statistically significantly within the traditional $p$-value framework.  The Bayesian analysis, however, revealed that around $90\%$ of the time one treatment option is superior to the other.

The BAT framework would allow the agency to take advantage of the information emerging from the trial, e.g. regarding the strength and the variability of the treatment effect.  BATs would enable the agency to adjust the trial in response to the incoming data, recruiting more participants as required, discontinuing ineffective treatment arms, allocating a greater share of incoming participants to the treatments likely to be most effective, or identifying the most responsive cohorts \citep{offer2021adaptive}.  In the limit of full adaptability, where the trial is adjusted after each individual data point is observed, and under standard assumptions, BATs lead to maximum information gain and benefits to the recruited trial participants at each point in the trial.   By contrast, fixed-design RCTs are based on information available prior to the trial, and they proceed, based on this information, without adjustment to the end of the trial.  Because they ignore incoming data, RCTs are less data-efficient.   Given the same sample size and similar data collection and monitoring, BATs extract more information and/or deliver more benefits to the participants, \citep{muller2006bayesian}  (see also Section \ref{sec:ethics}).  This gain in efficiency is clearly shown in Section~\ref{sec:realexample}.

Additionally, within the BAT framework, the agency could leverage prior evidence to augment the results of the trial. For instance, in our example, compelling prior evidence led the agency to test the treatment in the local context. Unlike traditional RCT-based frameworks that discard such evidence, the BAT approach integrates it with new data.\footnote{BATs provide two channels to incorporate information from prior studies:  Via the prior and via the model.  For example, hierarchical modeling enables borrowing of estimative strengths across related but independent experiments}    With strong prior evidence, a smaller, less costly BAT is sufficient to match the confidence level of a larger trial without it.

\subsection{Alignment with Decision-Making}
\label{sec:alignment}

The right-fit evidence system must support decision-makers with evidence-based insights on a range of decisions, such as discontinuing inefficient programs, expanding successful ones, trialing promising innovations, adjusting program parameters, and choosing between different program options (\cite{gertler2016impact,gugerty2018goldilocks},\cite{gugerty2018ten}).   The critical evidence-based decisions in the public (or private) sector are ultimately about how to allocate scarce public resources most effectively among the available alternatives \citep{mitchell2020instrumental}. The question about a treatment or support is never whether it works in the abstract, but whether it is worth allocating scarce resources to one vs. another program – for this cohort, in this location, in combination with other treatments.  To help answer this question, the evidence-based system must enable rigorous comparison of the efficacy of potential treatments relative to their estimated costs, particularly in the context in which these treatments would be applied.  

Traditional impact evaluation methodologies and evidence systems meet these criteria partially and leave significant gaps.  Take, for instance, the need for rigorous direct comparisons between treatments - a task, with which traditional methods have significant difficulties.   RCTs famously test the $p$-value of the treatment effect, widely criticized in the statistical community as misleading in decision-making \citep{wasserstein2016asa}.  The $p$-value framework does not answer the direct question of whether a specific treatment effect exists and, under most practical circumstances, is not suitable for deciding whether one treatment is better than another treatment \citep{lammers2023use,berry2010bayesian}.  

Traditional impact evaluation frameworks, focusing on rigorous assessment of treatment effects, often neglect implementation cost estimation, a key factor in resource allocation. Decision-makers require a clear understanding of the benefit-to-cost ratio, which balances treatment effects against costs.  Traditional RCTs prioritize establishing treatment effects' causality but handle cost estimation less rigorously. This skewed focus on the benefit side of the ratio results in uncertainty about the crucial metric of treatment effect per dollar spent, despite the significant effort expended on proving statistical significance.

More generally, traditional impact evaluation techniques suffer from highly varied levels of rigour along the decision-making logic chain.   Rigorous traditional evaluations, particularly those based on RCTs, are effective only in a limited phase of a program's lifecycle --  post-design but pre-rollout to the target population.  \citep{gugerty2018ten}.  Outside this phase, organisations perform ad hoc analyses.   In the evaluation design phase, ad hoc approaches result in underpowered evaluations \citep{gugerty2018goldilocks}, unable to provide reliable causal inference.  During the rollout phase, they lead to loss of information from pilot evaluations. Less rigorous analyses surrounding the more rigorous impact evaluations dilute the power of the rigorous methods.  This manifests in external validity issues \citep{williams2020external} and failures in scaling up successful pilots \citep[see, e.g.,][and references therein]{epstein2012program}.

The decision-theoretic BATs framework provides a flexible scaffold that can seamlessly \citep{bothwell2018adaptive} extend the rigour of the formal trial through the entire lifecycle of a program.   Unlike traditional impact evaluations, BATs can be advantageous at nearly any point:  in the early design phase when features are selected, in the trial phase where effectiveness for different cohorts is evaluated, and in the scaling phase where the most effective scaling pathway is designed – without loss of rigour or information at handoffs.  

BATs promote decision-making discipline by making explicit the assumptions often implicit in traditional impact evaluations.  The construction of priors (Section~\ref{sec:bats}) encourages rigorous evaluation of available existing evidence.  The selection of a model (Section~\ref{sec:bats}) promotes a thoughtful examination of the theory of change, including the alignment of observable outcomes with the objectives of the organisation.  It also promotes the identification of the critical unknowns in the theory of change, methods to test these, and their influence on the likelihood of successfully scaling up a pilot or initiative.

The BAT framework is capable of integrating data from multiple sources to extract information.  For example, monitoring data provides highly valuable information about the local context, such as selective take-up, heterogeneity of impacts, or spillover effects, not captured in traditional trials (\cite{ravallion2009should},\cite{ravallion2020should}). This information is vital for gauging how treatments proven elsewhere might apply to the local context.  Impact evaluations of past initiatives -- including from unsuccessful pilots and underpowered RCTs -- represent particularly valuable sources of local information.  Within the Bayesian framework, every experiment -- even one that's imperfect -- yields information, and it is possible to use modern tools to capture this information with scientific rigour.  The key is to recognize that that inferences from different methodologies vary in confidence and bias, and account for these variations.

BATs focus directly on evaluating the range of probable outcomes of a treatment for a target population, given all the available information, enabling direct comparisons between treatments (see also Section \ref{sec:cost}). BATs are also well suited for testing multiple treatment combinations and for identifying the most responsive cohorts \citep{juszczak2019reporting}, an increasingly crucial capability given the rapid growth in candidate interventions \citep{leigh2009evidence}.  This feature is particularly relevant in social policy, where treatments are often delivered in combination.   Lastly, the BATs decision-theoretic framework can put cost and treatment effect estimation on an equal footing, ensuring that both are estimated with commensurate precision \citep{berry2010bayesian}.

\subsection{Ethics and Beneficence}
\label{sec:ethics}

Bayesian adaptive trial designs offer several advantages and, on balance, ameliorate ethical concerns surrounding traditional fixed trial designs \citep{legocki2015clinical}, particularly when applied to social policy contexts.  BAT approaches lead to faster results; they balance the information gain for future cohorts with benefits to current participants; they promote equipoise in the face of strong priors; and they provide clear channels for community input in trial design.

BAT's efficiency results in smaller required sample sizes and faster access to better treatments, which benefits both current participants and future target cohorts.  Additionally, adaptive designs offer trial participants a higher chance of receiving the most effective treatment and exposing fewer participants to ineffective interventions.  Critics of BATs' ethical superiority highlight the ethical dilemmas posed by later participants receiving more effective treatments than earlier ones. This raises questions about justice and informed consent \citep{saxman2015ethical,michiels2016statistical}. However, advocates of adaptive designs counter that fixed designs often prioritize future benefits over current participants, favoring a ``collective ethics'' approach \citep[see, e.g.,][]{palmer1999ethics,pullman2001adaptive}.

In the context of social policy, commentators observe that evaluative RCTs often breach the principle of equipoise, as sponsors increasingly demand that any RCT proposal be ``backed by highly-promising prior evidence, suggesting it could produce sizable impacts on outcomes…'' (Laura and John Arnold Foundation 2018, quoted in \cite{bedecarrats2020randomized}).  When existing research strongly indicates likely treatment effects, a randomized design ignoring this evidence effectively deprives control group participants of treatment. The BAT framework, by explicitly constructing and incorporating the prior into its trial design, partially restores the condition of equipoise. 

Importantly, in the social policy context, the BAT framework promotes the close involvement of the community and on-the-ground staff in experiment design -- through explicit elicitation of priors and theories of change.  The co-design of treatments, interventions, and trial designs with the community is critical to meeting the requirement of beneficence \citep{deaton2020randomization}.

\subsection{Contextualisation and Personalisation}

One of the critical problems in evidence-based policy-making is how to best apply the evidence from the growing number of rigorous impact evaluations worldwide in the local context \citep{williams2020external,leviton2017interaction}. \cite{williams2020external} argues that failures in external validity often stem from the interplay between a policy's theory of change and a dimension of implementation context. Policymakers must navigate a trade-off between the robustness and relevance of evidence from other contexts and their understanding of local conditions. They face choices: to replicate, modify, or create new approaches. The BAT framework, through its construction of priors and inference models, offers a structured method to organize and analyze the information needed for these decisions.  

High-quality RCTs, especially systematic reviews and meta-analyses of similar RCTs, reveal more universal, context-independent responses to treatments, producing transportable results \citep{williams2020external}. For treatments with such robust evidence, local decision-makers must determine the most beneficial strategies for their specific cohort: identifying who benefits most, the optimal treatment dosage, the best combination of treatments for each sub-group, and defining these relevant sub-groups. BATs offer an effective framework for addressing these critical questions.

Governments can improve the effectiveness and efficiency of social programs by adapting policies to local contexts and the needs of individuals.    Contextualisation in social policy considers the local socio-economic and cultural context affecting the implementation and outcomes of policies.  Personalisation refers to tailoring services and supports to the unique needs and circumstances of individuals. Technology has enabled personalisation and contextualisation in the private sector, as manifested, for example, in recommendation engines and marketing campaigns \citep{grewal2011innovations,schrage2021transformational,mostaghel2022digitalization}.  Bayesian methods are increasingly popular for personalisation and contextualisation in the private sector as they enable the continuing balancing of exploiting existing information and acquiring data to continue learning in order to, e.g., adapt to changing contexts and preferences \citep{rendle2012bpr}.

\subsection{Validating Theories of Change}
\label{sec:Theoryofchange}

A feature of BATs that proponents of RCTs view as a weakness is that Bayesian inference depends on a probabilistic model as discussed in Section~\ref{sec:bats}– effectively a quantified theory of change – whereas RCTs rely solely on the assumption of independence of individual observations.   For social policy applications, the ability of the Bayesian framework to incorporate a theory of change is a significant benefit.   It enables not only rigorous sharing of information across trials, e.g., trials of the same treatment across different locations or cohorts, but also, given enough data, the evaluation and validation of competing theories of change. In complex settings, such as social policy or environmental conservation, a number of theories of change may be plausible.  The BAT framework enables the incorporation of a mixture of competing theories of change for inference.   As informative data are collected, the Bayesian framework will point to some of the theories being more likely \emph{given the data} than others.  It can also accommodate mechanism experiments, such as those of \cite{ludwig2011mechanism}.   That is, in addition to the question “Does it work?” answered by classic RCTs, the use of the Bayesian framework can help us answer the question “Why does it work?” and enable us to discover high-value treatments to test.

\section{Numerical Example: Bayesian adaptive learning of the relation between student performance and external tutoring hours}
\label{sec:realexample}

In this section, we provide a simulated example to demonstrate the data efficiency of BATs.  We simulate a Bayesian adaptive trial to find the optimal number of tutoring hours delivered (per student per month) in order to lift students' educational outcomes.  
In this illustrative example, collecting new data is costly. Nonetheless, by using an adaptive trial framework we can substantially reduce the required observation sample size to reach a required level of accuracy. 

The overarching goal is to learn the function that maps the number of tutoring hours per month, $x$, to the increase in students' performance, $y$, as efficiently as possible.  There are two features of this function we wish to learn - the number of tutoring hours per month which will maximize;
\begin{enumerate}[(A)]
    \item  Expected student performance.
    \item Expected rate of increase in the student's performance. 
\end{enumerate}
In addition, we wish to do this as efficiently as possible so that the {\it experimental dose} of tutoring hours must be chosen in an order that will maximally learn (reduce uncertainty) about the function $\theta$ where it is likely to find an optimal trade-off between objectives (1) and (2).  These objectives are formalised as a utility function in Section~\ref{eq.utility.def}.


\subsection{Formal setting}

\begin{figure}[h]
\includegraphics[width=8cm]{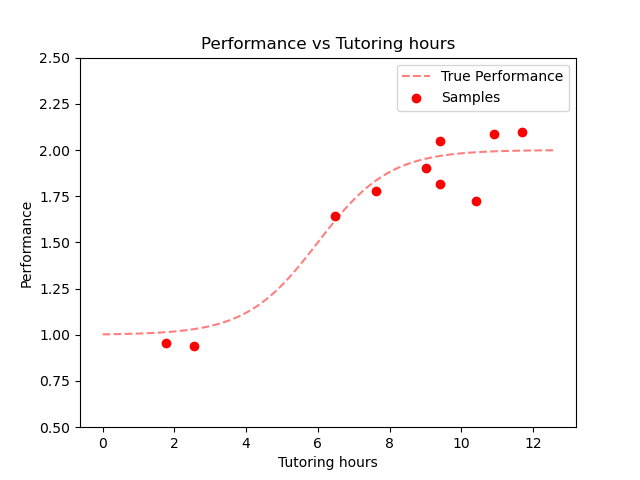}
\centering
\caption{10 Samples drawn from the model defined by
\eqref{eq.f.star} and \eqref{eq.f.example} where 
$m=6$, $b=1$ and $\sigma_\epsilon = 0.1$. The true performance, $\theta(x)$, is represented by the dashed line.}
\label{fig.response}
\end{figure}

Let the true \emph{students' performance}, $\theta$, be a function of tutoring hours per month, $x$, and be in the form of the following logistic function:
\begin{align}
    \theta(x) = \frac{1}{1+\exp(-x+m)} + b,
\label{eq.f.star}
\end{align}
with midpoint, $m=6$ and intercept, $b=1$. 
We assume that observations of students' learning  $y$, given tutoring hours, $x$, are a noisy signal around this true function. Specifically, we have
\begin{align}
    y = \theta(x) + \epsilon, \qquad \text{ where } \epsilon \sim \mathcal{N}(0, \sigma_{\epsilon}^2).
\label{eq.f.example}
\end{align}
In the above equation, $\mathcal{N}(0, \sigma_{\epsilon}^2)$, denotes a normal distribution with mean $0$ and standard deviation $\sigma_{\epsilon}$ which we set to be $\sigma_\epsilon = 0.1$.
Figure~\ref{fig.response} plots the true performance, $\theta(x)$, (dashed red line) along with 10 randomly sampled observations, $y$, drawn from this illustrative model.  

We wish to design a trial to learn the function $\theta(x)$, and specifically the features (A) and (B) given above,  given some data. In a Bayesian context a natural estimate of this function is its posterior mean, denoted by $\hat{\theta}$, and to obtain this estimate we begin by placing a prior over this function.  In doing so we need to ask what
type of prior information we need to impose on a function, so that, given
some data, an estimate of this function will have desirable properties.   These desirable properties are that, 
\begin{enumerate}
    \item The estimate of function is relatively smooth,
    \item The prior be flexible enough to admit a large range of estimated functions. 
\end{enumerate}
One such prior is a Gaussian Process prior, $\mathcal{GP}$, \citep{williams2006gaussian} so that we have
\begin{align}
    p(\theta) = \mathcal{GP}(\theta; \mu_, \Omega),
    \label{eq.gp.prior}
\end{align}
where the mean $\mu=0$ and the covariance matrix, $\Omega$, is chosen to be a \emph{Radial basis function} (RBF) with length scale 2.0, \cite{wood13}.  This prior ensures smoothness but does not impose many other restrictions, for example it does not suppose that the function is monotonic.

Next, we need a likelihood function which connects observed outcomes $\mathbf y$ to $\theta(x)$.  This is given by is Equation~\eqref{eq.f.example} so that $\mathbf y|\theta(x)\sim N\left(\theta(x),I\sigma_{\epsilon}\right)$.  The task of a trial is to design an experiment that learns about $\theta(x)$ and its features as efficiently as possible by generating data $\mathbf y$ to update our prior belief $p(\theta)$.

\subsection{Two Trials}
We now consider two trials, both of size $n=12$. 
\subsubsection{Fixed Design}
In the first trial we fix the number of tutoring hours $\mathbf x=(x^{[1]},\ldots,x^{[12]})$ to be equally spaced between in the interval [0,12] and record the outcomes of student performance $\mathbf y=(y^{[1]},\ldots,y^{[12]})$, so that our data denoted by $\D$ is $\D^{[1:12]} = \left\{ (x^{[1]}, y^{[1]}), \ldots (x^{[12]}, y^{[12]}) \right\}$. 
 
\begin{figure}[h]
\includegraphics[width=8cm]{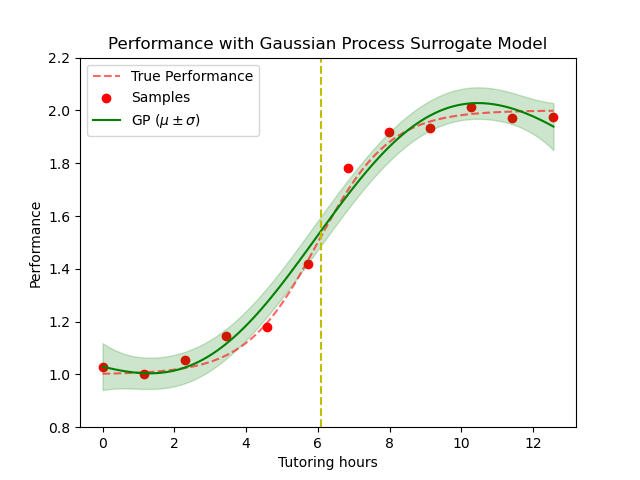}
\centering
\caption{Posterior Gaussian Process, $\mathcal{GP}(\theta; \mu_{\D^{[1:12]}}, \Omega_{\D^{[1:12]}})$ for the {\bf fixed design} trial, fitting the the students' performance, $\theta$, versus external tutoring hours ($x$) where the observations, $\D^{[1:12]}$, are depicted by filled circles; 
The red dashed line shows the true performance, $\theta(x)$, 
and the green line and shaded confidence interval represent
$\mu_{\D^{[1:12]}}(x) \pm 0.95 \sqrt{\Omega_{\D^{[1:12]}}(x,x)}$.
Yellow line: Tutoring hours where a combination of features (A) and (B) (formalised by \eqref{eq.u12}) is maximised. 
}
\label{fig.response.gp}
\end{figure}

 Figure~\ref{fig.response.gp}  shows a plot of this data together with an estimate of the function $\hat{\theta}=\mu_{\D^{[1:12]}}$ (solid green line) and the true function (dotted red line)  and the shaded credible intervals  $\mu_{\D^{[1:12]}}\pm 0.95 \cdot \sigma_{\D^{[1:12]}}(x)$ where 
\begin{align*}
    \sigma_{\D^{[1:12]}}(x) := \sqrt{\Omega_{\D^{[1:12]}}(x,x)-\Omega_{\D^{[1:12]}}(x,x)^2/(\Omega_{\D^{[1:12]}}(x,x)+\sigma_{\epsilon}^2)}.
\end{align*}

\subsubsection{Adaptive Design}

Another possible design is the following\\
{\bf Algorithm 1}
\begin{enumerate}
    \item Set $t=1$.
    \item Randomly draw $x^{[t=1]}$ uniformly from the interval [0,12].
    \item Observe $y^{[t]}$.
    \item Compute $\mu_{\D^{[1:t]}}(x)$ and $\sigma_{\D^{[1:t]}}(x)$ and  the utility function $U_3(x|\D^{[1:t]})$, given by Equation~\eqref{eq.u3}.
    \item Find $x^*=\arg\max_x U_3(x|\D^{[1:t]})$.
    \item Set $t=t+1$.
     \item Set $x^{[t]}=x^*$.
     \item Repeat Steps 3-7 until $t=12$.
\end{enumerate}

\begin{figure}
\begin{subfigure}{.5\textwidth}
  \centering
   \includegraphics[width=0.95\textwidth,trim=16mm 2mm 20mm 15mm, clip=true]{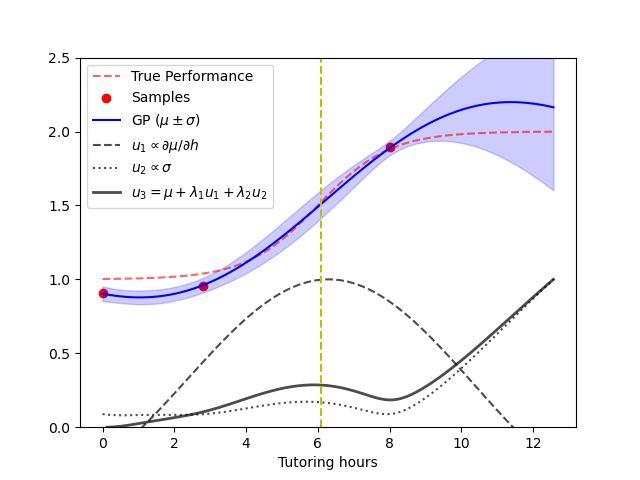}
  \caption{}
  \label{fig:sfig1}
\end{subfigure}%
\begin{subfigure}{.5\textwidth}
  \centering
  \includegraphics[width=0.95\textwidth,trim=16mm 2mm 20mm 15mm, clip=true]{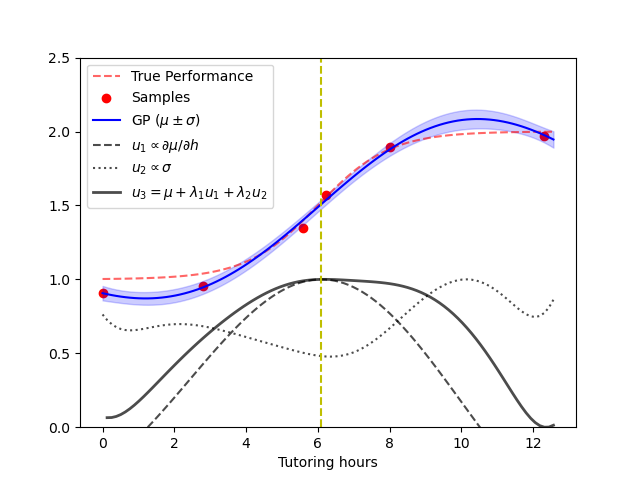}
  \caption{}
  \label{fig:sfig2}
\end{subfigure}
\\
\begin{subfigure}{.5\textwidth}
  \centering
  \includegraphics[width=0.95\textwidth,trim=16mm 2mm 20mm 15mm, clip=true]{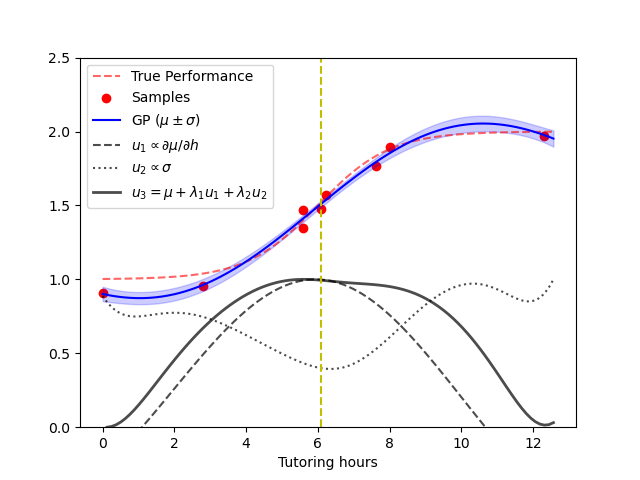}
  \caption{}
  \label{fig:sfig3}
\end{subfigure}%
\begin{subfigure}{.5\textwidth}
  \centering
  \includegraphics[width=0.95\textwidth,trim=16mm 2mm 20mm 15mm, clip=true]{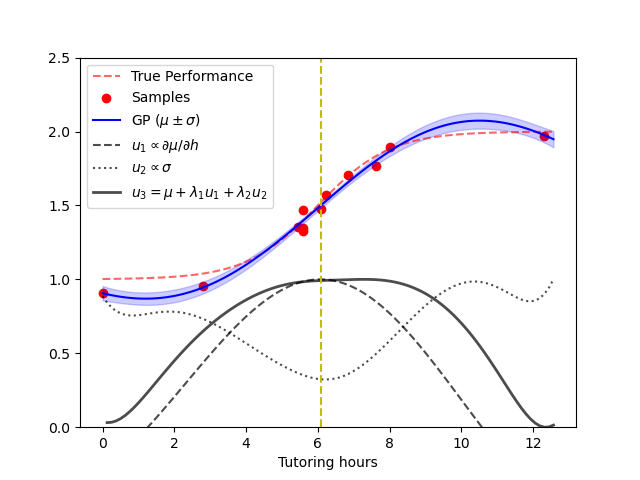}
  \caption{}
  \label{fig:sfig4}
\end{subfigure}
\caption{Bayesian adaptive trial (BAT) applied to model \eqref{eq.f.example}. Red dashed line: the true performance, $\theta(x)$, to be approximated (see equation \eqref{eq.f.star}). Circles, $\D^{[1:t]}$: data points; 
Blue line: GP's mean, $\mu_{\D^{[1:t]}}(x)$; 
Dashed black line: proportional to the slope of GP's mean, $\frac{\partial \mu_{\D^{[1:t]}}(x)}{\partial x}$; 
Dotted line: proportional to GP's standard deviation, $\sigma_{\D^{[1:t]}}(x)$; 
Solid back line represents the acquisition function~\eqref{eq.u3} with $\lambda_1=30$ and $\lambda_2=10$. 
Yellow line: Tutoring hours where a combination of features (A) and (B) (formalised by \eqref{eq.u12}) is maximised. 
}
\label{fig:adaptive}
\end{figure}

Figure~\ref{fig:adaptive} panels $(a)$, $(b)$ $(c)$ and $(d)$ show estimates of $\mu_{\D^{[1:t]}}(x)$ (blue solid lines) and corresponding 95\% credible intervals (blue shaded area) after applying Algorithm 1 for  $t \in \{3, 6, 9, 12\}$ respectively.  
As before, 
the true performance, $\theta(x)$, (that should be approximated by GP) is plotted by the red dashed line; the filled circles represent the existing data, $\D^{[1:t]}$.

In addition Figure~\ref{fig:adaptive} panels $(a)$, $(b)$ $(c)$ and $(d)$ show the utility functions, $U_1(x|\D^{[1:t]})$, $U_2(x|\D^{[1:t]})$, and $U_3(x|\D^{[1:t]})$ for $t \in \{3, 6, 9, 12\}$. The dashed black line, $U_1$, is proportional to the slope of the expected performance curve vs tutoring hours. As such the pick of this curve corresponds to the tutoring time, $x^*_{\text{II}}$, where the effect of tutoring is maximal
(see equation~\ref{eq.ii}). 
Similarly, the dotted black line, $U_2$ is proportional to the GP's posterior standard deviation and maximises at the tutoring time, $x^*_{\text{II}}$, where the relation between the tutoring hours and performance is the most uncertain (see equation~\ref{eq.iii}).
The solid black curve, $u_3$, represents a utility function that is a linear combination of objectives I to III, that is:
\begin{align}
    U_3(x | \D^{[1:t]}) = 
    \mu_{\D^{[1:t]}}(x) +  
        \lambda_1 \cdot \frac{\partial \mu_{\D^{[1:t]}}(x)}{\partial x}  +
        \lambda_2 \cdot \sigma_{\D^{[1:t]}}(x), 
    \label{eq.u3}
\end{align}
with combination weights chosen to be $\lambda_1=30$ and $\lambda_2=10$.

It can be seen that only after a few draws, the overall utility function, $U_3$, is maximised in the adjacency of $x=6$h, where the utility function \eqref{eq.u3} is maximised.

\begin{figure}
\begin{subfigure}{.5\textwidth}
  \centering
   \includegraphics[width=0.95\textwidth,trim=16mm 2mm 20mm 15mm, clip=true]{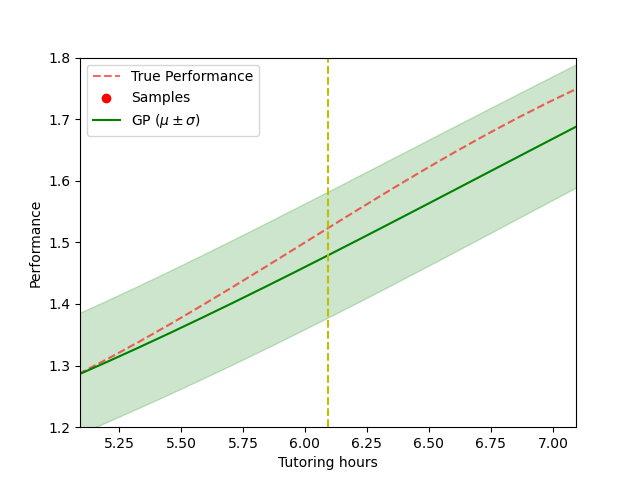}
  \caption{}
  \label{fig:ssfig1}
\end{subfigure}%
\begin{subfigure}{.5\textwidth}
  \centering
  \includegraphics[width=0.95\textwidth,trim=16mm 2mm 20mm 15mm, clip=true]{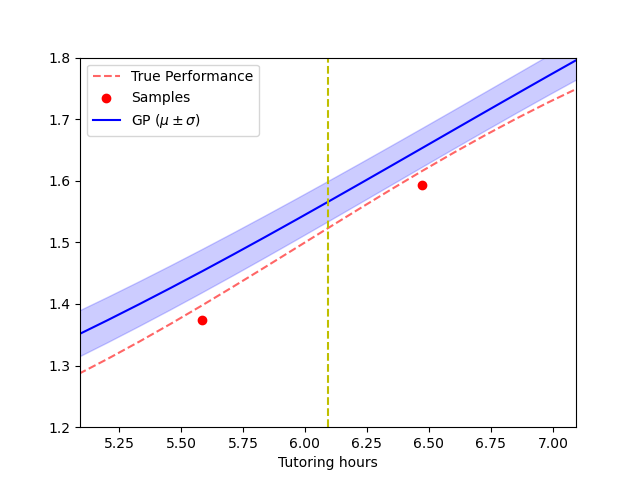}
  \caption{}
  \label{fig:ssfig2}
\end{subfigure}
\\
\begin{subfigure}{.5\textwidth}
  \centering
   \includegraphics[width=0.95\textwidth,trim=16mm 2mm 20mm 15mm, clip=true]{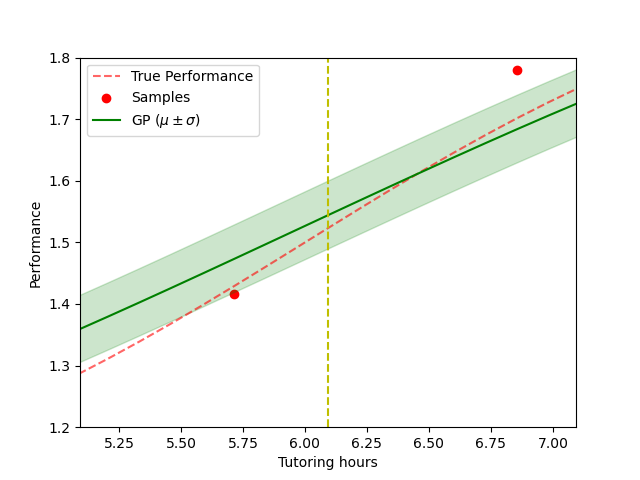}
  \caption{}
  \label{fig:ssfig3}
\end{subfigure}%
\begin{subfigure}{.5\textwidth}
  \centering
  \includegraphics[width=0.95\textwidth,trim=16mm 2mm 20mm 15mm, clip=true]{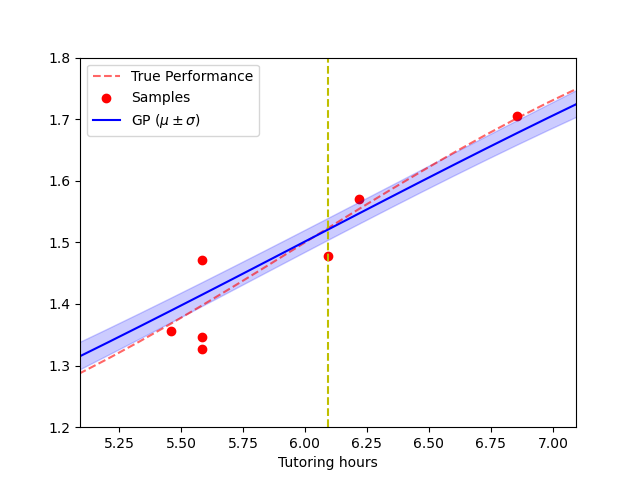}
  \caption{}
  \label{fig:ssfig4}
\end{subfigure}
\caption{
Bayesian Adaptive trial (BAT) (left column) versus Fixed design (right column). 
Red dashed line: the true performance, $\theta(x)$, to be approximated. 
Circles, $\D^{[1:t]}$: data points; 
Blue and Green lines: GP's mean, $\mu_{\D^{[1:t]}}(x)$;  
Yellow vertical line: Tutoring hours at which the linear combination of the performance and effect of tutoring is maximised (see \eqref{eq.u12}).  
First row: 6 data points collected from the interval [0,12] (a subset of which are in the interval [5,7]) (a) By fixed design, (b) by BAT.
Second column: 12 data points collected from the interval [0,12]  (a) By fixed design, (b) by BAT.
}
\label{fig:comparison}
\end{figure}
\subsubsection{Comparison of Trials}
Figure~\ref{fig:comparison} panels (a) and (b) compare the performance of the fixed design (a) and the adaptive design (b) for $t=6$ while panels (c) and (d) are similar plots for $t=12$.  Figure~\ref{fig:comparison} demonstrates the superiority of the adaptive design both in terms of accuracy and efficiency.  For both $t=6$ and $t=12$, the estimated function from the adaptive trial design in the vicinity of the optimal rate of learning is closer to the true function than the estimate from the fixed trial design. Additionally, the uncertainty in the estimate for the adaptive trial design for $t=6$ is less than the uncertainty in the fixed design for $n=12$.  This shows that by being adaptive, and specific about the quantities we wish to learn, we get better results in the adaptive trial design with half of the samples needed for the fixed trial design.   

We now provide details on the construction of the utility functions.
\subsection{Utility function}
\label{eq.utility.def}
In this example, we consider 3 different utility functions which, for  
ease of exposition, only depend on the action, $x$, and performance, $\theta$. As mentioned earlier, in this illustrative example, the utility function is a linear combination of three objectives: 
\vspace{1mm}
\\
\textbf{I. Maximising the students' performance.}
Clearly, the prime goal of providing tutoring hours, $x$ (per student per month), is to maximise students' expected performance. In the mathematical formalism, this is equivalent to finding a tutoring hour, $x^*_{\text{I}}$, such that:
\begin{align*}
    x^*_{\text{I}} = \argmax \, \mu_{\D^{[1:t]}}(x), 
\end{align*}
since given the existing data, $\D^{[1:t]}$, the expected performance is approximated by the GP's (posterior) mean, $\mu_{\D^{[1:t]}}(x)$.

Note that according to the presented model, 
increasing the tutoring hours, $x$, (per student per month)
monotonically increases the expected student's performance.
However, providing tutoring hours is costly and given the limited budget, assigning boundless tutoring hours per student is not feasible. This leads to a second objective:
\vspace{1mm}
\\
\textbf{II. Maximising the effect of tutoring.}
Given the limited resources, it is desirable to find tutoring hours that  
lead to a maximal increase in the student's performance. 
This is equivalent to finding a point, $x^*_{\text{II}}$, where the slope of the performance vs tutoring curve is maximal. That is: 
\begin{align}
    x^*_{\text{II}} = \argmax
    \frac{\partial \mu_{\D^{[1:t]}}(x)}{\partial x} .
    \label{eq.ii}
\end{align}
Similar utility functions have been demonstrated to succeed in robotics for the exploration of physical terrain, focusing on those areas with maximal slope \citep{Morere2017}.

Note that objectives (A) and (B) that are introduced in the introduction of this section are obtained if a linear combination of I and II is maximised. This corresponds to the following utility function:
\begin{align}
    U_{1,2}(x | \D^{[1:t]}) = 
    \mu_{\D^{[1:t]}}(x) +  
        \lambda_1 \cdot \frac{\partial \mu_{\D^{[1:t]}}(x)}{\partial x},  
    \label{eq.u12}
\end{align}
where we choose $\lambda_1=30$.

Objectives I and II are both based on the assumption 
that given the existing data, $\D^{[1:t]}$, the expected performance versus tutoring, $\mathbb{E}[\theta(x)]$, is approximated sufficiently well by GP's mean, $\mu_{\D^{[1:t]}}(x)$. That is:
\begin{align*}
    \mu_{\D^{[1:t]}}(x) \approx \mathbb{E}[\theta(x)].
\end{align*}
Otherwise stated, objectives I and II are two different kinds of \emph{exploitation} (of the already gathered data $\D^{[1:t]}$) and can only be attained if the relation between the performance and tutoring is established by sufficient \emph{exploration}. This leads to a third objective:  
\vspace{1mm}
\\
\textbf{III. Minimising the uncertainty in the relation between tutoring hours and performance.}
The uncertainty in the curve fitting is reflected in GP's posterior standard deviation,    
$\sigma_{\D^{[1:t]}}(x) := \sqrt{K_{\D^{[1:t]}}(x,x)}$.
As such, a pure explorative objective can be defined by gathering new data from a setting that is associated with maximal uncertainty, this is, the tutorial hour (per month), $x^*_{\text{III}}$, where GP's standard deviation is maximal:
\begin{align}
    x^*_{\text{III}} = \argmax \sigma_{\D^{[1:t]}}(x).
    \label{eq.iii}
\end{align}

Putting everything together, we generate the next data-point, $(x^{[t+1]}, y^{[t+1]})$,  by assessing students' performance after $x^{[t+1]}$ hours of monthly training where $x^{[t+1]}$ is the quantity that maximises a linear combination of the introduced objectives I to III:  
\begin{align}
    x^{[t+1]} = \argmax \left( 
        \mu_{\D^{[1:t]}}(x) +  
        \lambda_1 \cdot \frac{\partial \mu_{\D^{[1:t]}}(x)}{\partial x}  +
        \lambda_2 \cdot \sigma_{\D^{[1:t]}}(x)      
    \right),
    \label{eq:adaptive.total}
\end{align}
where $\lambda_1$ and $\lambda_2$ are tunable parameters that determine the trade-off between the three objectives.

\section{BATs and Organisational Learning}
\label{sec:learning}

Bayesian Adaptive Trials (BATs) represent not just an efficient experimental methodology, but also a foundational shift towards a right-fit evidence system. The most profound benefit of incorporating BATs lies in the cultural transformation it fosters – a shift from a binary perspective of ``success and failure'' to a mindset centered on continuous learning and improvement.  Moving away from binary answers to “Does it work?” to probabilistic thinking would ameliorate the agency problems that stand as a barrier to broader adoption of impact evaluations.

There is often a reluctance among political figures to engage in rigorous evaluations of initiatives.  One explanation is that in public sector decision-making, there exists a pervasive illusion of control, and policymakers often commit to delivering results through initiatives with uncertain outcomes. Not surprisingly many initiatives do not achieve their intended results, and therefore many evaluations report more shortcomings than successes.

This prevailing reluctance necessitates a paradigm shift. The most effective way to learn what works, and under what circumstances, is through embedding empirical experimentation within the contexts where initiatives are intended to operate. Public sector initiatives inherently serve as such experiments. By embracing this learning perspective, the need for public sector decision-makers to predicate their careers on the success of inherently uncertain programs can be significantly mitigated. 

A pivot towards a culture that values empirical experimentation and learning from each initiative, regardless of its outcome has other advantages. It promotes more efficient utilisation of resources, as it acknowledges and leverages the valuable insights gained from the experiences of these programs, rather than disregarding information if the evaluations are not positive, and also contributes to a more responsible and sustainable approach to governance.

Advancements in digital technology such as the creation of service delivery channels that facilitate low-cost experimentation and data collection are progressively enabling the practical implementation of adopting a BAT mindset for organisational learning. These innovative channels, such as online learning platforms, can support the implementation of ``perpetual trials'' and foster ongoing learning loops. When one compares the low cost of these learning options with the inherent constraints and significant costs of human resources and organisational capacities required for the more traditional evaluation methods it is impossible to escape the conclusion that a more focused and agile approach may be advantageous.  Such developments hold the promise of making continuous experimentation and data gathering more feasible and less burdensome.

\section{Conclusions}

This paper underscores the transformative potential of the BAT framework as an impactful and efficient methodology for social policy impact evaluations and, more broadly, right-fit evidence systems. By transcending the limitations of traditional evaluation methods like RCTs, BATs offer a dynamic, cost-effective, and highly adaptable approach that is in line with the policy cycle. The ability of BATs to integrate various data types through Bayesian reasoning, while continuously adapting to new information, makes them well-suited for real-world policy applications where complexity and variability are the norms. This approach aligns with the need for more agile and rigorous methods in policy evaluation, especially in contexts where decisions must be made swiftly and with limited resources. The success of BATs in fields like medicine and marketing further underscores their potential in the realm of social policy. 

As technological and algorithmic advancements continue to evolve, the implementation of BATs in social policy could mark a significant shift towards more effective, evidence-based decision-making. This shift is not just about adopting a new methodological framework; it represents a broader movement towards a culture of continuous learning and adaptive management in the pursuit of social good. Ultimately, BATs stand to greatly enhance our ability to discern the most impactful and beneficial policy interventions, thus fostering a more informed, efficient, and responsive approach to social policy development and implementation.

\section*{Funding}
This research was supported by the Paul Ramsay Foundation under the \emph{Thrive: Finishing School Well} project.

\section*{Competing Interests}
The authors declare no competing interests.  Although B.G. is an employee of the funding organization, the Paul Ramsay Foundation, he has not been involved in decision-making processes related to the allocation or management of grant funding, thus mitigating a perceived conflict of interest.

\section*{Data Availability}
The paper illustrated the conceptual framework using simulated data, as described in Section~\ref{sec:realexample}.

\section*{Author Contributions}
Conceptualization: all authors; Data curation: N/A; Formal analysis: S.C., A.L., and H.M.A.; Funding acquisition: S.C and R.M.; Investigation: all authors; Methodology: all authors; Project administration: S.C. and G.F.;  Software: H.M.A. and R.M.; Supervision: S.C.; Validation: S.C., A.L., and H.M.A.; Visualization: S.C. and H.M.A.; Writing—original draft: S.C., A.L, and H.M.A; Writing—review and editing: all authors.

\bibliographystyle{apalike}

\end{document}